\documentclass{elsart}

\usepackage[]{graphicx}
\usepackage{amssymb}

\begin{document}

\begin{frontmatter}

\title{Exact diagonalization study of the spin-1 two-dimensional $J_1$-$J_3$  Heisenberg model on a triangular lattice}

\author{P.~Rubin$^\dagger$\thanksref{mail}},
\author{A.~Sherman$^\dagger$}

\address{$^\dagger$Institute of Physics, University of Tartu,
Ravila 14c, 50411 Tartu, Estonia}
\thanks[mail]{Corresponding author:
E-mail: rubin@fi.tartu.ee}

\begin{abstract}
The spin-1 Heisenberg model on a triangular lattice with the ferromagnetic nearest- and antiferromagnetic third-nearest-neighbor exchange interactions, $J_1=-(1-p)J$ and $J_2=pJ$, $J>0$ $ (0 \leq p \leq 1)$, is studied with the use of the SPINPACK code. This model is  applicable for the description of the magnetic properties of NiGa$_2$S$_4$. The ground,  low-lying excited state energies  and  spin-spin correlation functions have been found for  lattices with N=16 and N=20 sites with the periodic boundary conditions.
These results are in qualitative agreement with  earlier authors' results obtained with Mori's  projection operator technique.

\end{abstract}


\end{frontmatter}

The spin-1 $J_1$-$J_3$  Heisenberg model on a triangular lattice, which takes into account the ferromagnetic nearest- and antiferromagnetic third-nearest-neighbor exchange interactions ($J_1$ and $J_3$, respectively) is of
interest as a minimal model for the description of magnetic properties of the compound NiGa$_2$S$_4$ \cite{SNakatsuju}. These properties are mainly determined by the two-dimensional triangular lattice of Ni$^{2+}$ ions with the spin $S=1$. In particular,
the magnetic neutron scattering experiment revealed the incommensurate short-range order \cite{SNakatsuju} - the scattering intensity had a maximum at some incommensurate vector $\bf Q_{exp}$.
The  classical version of the $J_1$-$J_3$  model  was proposed in Ref. \cite{SNakatsuju}, authors of which
 were able to reproduce the observed  incommensurate order with the vector $\bf Q_{exp}$ by fitting the ratio $J_1/J_3$.
  The quantum $J_1$-$J_3$  Heisenberg model was investigated in our recent papers \cite{r1, r2} with the use of Mori's  projection operator technique  \cite{Mori}.
  It was shown that at zero temperature, depending on the ratio $J_1$/$J_3$, the system is characterized by the ferromagnetic ordering, spin disorder, incommensurate and commensurate antiferromagnetic ordering. At $J_1/J_3 \approx -0.22$  the model describes key features observed [1] in NiGa2S4 - the incommensurate  antiferromagnetic
  short-range order at finite temperature, the quadratic temperature dependence of specific heat and the shape of the uniform susceptibility.

 Applying Mori's method one has to use a number of approximations. Therefore, it is of interest to study the same model with  another  method and compare
 obtained results. In this work we employ Schulenburg's SPINPACK code \cite{spin}.  This package is dedicated for exact diagonalization (ED) of finite spin system  using Lanczos algorithm.

The Hamiltonian of the  model  reads
\begin{equation}\label{hamiltonian}
H=\frac{1}{2}\sum_{\bf nm}J_{\bf nm}\left(s^z_{\bf n}s^z_{\bf
m}+s^{+1}_{\bf n}s^{-1}_{\bf m}\right),
\end{equation}
where $s^z_{\bf n}$ and $s^\sigma_{\bf n}$ are the components of the
spin-1 operators ${\bf s_n}$, {\bf n} and {\bf m} label sites of the triangular lattice, $\sigma=\pm 1$.
As mentioned above, we take into account the nearest-neighbor and third-nearest-neighbor interactions,  $J_{\bf nm}=J_1\sum_{\bf a}\delta_{\bf n,m+a} + J_3\sum_{\bf A}\delta_{\bf n,m+A} $ with the vectors {\bf a}  and {\bf A=2a} connecting the respective  sites.
Here the frustration parameter $p$ is introduced, $J_1=-(1-p)J$, $J_3=pJ$. $J > 0$ will be used below as the unit of energy.

 For  $ 0 \leq p \leq 1$ we  found energies of the ground, low-lying excited states and spin-spin correlation functions for lattices containing $N= 16$ and $20$ sites. These lattices are shown in Fig.~1. The periodic boundary conditions were used.
From our earlier study of the quantum  $J_1$-$J_3$  Heisenberg model on a $216 \times 216 $ lattice with Mori's method \cite{r1, r2} it is known that the system is ferromagnetically ordered in the interval
$0 < p < p_{cr}$, $p_{cr} \approx 0.2$, when the ferromagnetic coupling  $|J_1|$ is larger than $J_3$. We have found the same value also in the classical $J_1$-$J_3$   model on an infinite lattice.

As seen from  Figs. 2 and 3, the ground state (GS) of the N-site lattice is transformed
from the classic ferromagnet (S=N) to the singlet state at some critical value of
the frustration parameter $p$.
For the lattices with N=16 and N=20 these critical values
are $p_{16} \approx 0.45$ and $p_{20} \approx 0.28$, respectively.
 It can be supposed that with the rise of the lattice size this critical
value will tend to the value $ p_{cr}$ obtained in \cite{r1}.
According to \cite{r1} a transition from the ferromagnetically ordered state to a spin disorder occurs at this value of the frustration parameter $ p_{cr}$.
 In Figs. 2a and 3a the dependencies of the GS energy ($\textmd{E}_{{GS}}$, S=N) and the first excited state ($\textmd{E}_{{1}}$, S=15 and S=19, correspondingly) on $p$ are presented. These dependencies  are linear. Differences between the GS and excited-state energies disappear
at the critical values $p_{16}$ and    $p_{20}$. For $p > p_{16}$ and $p > p_{20}$ the energies of the low-lying
 states  are shown in Figs.~2b and 3b. The GS is characterized by S=0. The lowest excited states are characterized by
S=0 ($E_2$) and S=1 ($E_3$). Notice that relative positions of the curves for the first singlet $E_2$ and the first triplet $E_3$ excitations are different
for N=16 and N=20. Apparently the difference is related to the small size  of the  lattices and difference in their shapes.

The spin gap - the
energy difference between the first excited triplet   and the singlet ground state - is shown in  Fig. 4 for N=16 and  N=20.

In  Fig.~5 we compare the dependencies  of the GS energy per site   on the frustration parameter $p$ in the N=20,  N=16 lattices, obtained by  ED, and
in a $216 \times 216 $ lattice
obtained by Mori's  projection operator technique \cite{r1}. As a whole these dependencies  are similar. As one can see, all these plots are linear in the ferromagnetic region ($p < p_{16}, \, \,  p_{20}$ or $p_{cr}$). Besides, $ p_{cr} < p_{20} < p_{16}$. This sequence of the critical values of  $p$ seems reasonable because  Mori's result was  obtained in the largest  lattices.  For $p > p_{16},  \, \, p_{20}$ or $p_{cr}$ all curves have  maxima. However, positions of the maxima are different:  $p \approx 0.4$ in the results obtained with Mori's technique,
$p \approx 0.7$ for N=16 and $p \approx 0.9$ for N=20.

Spin correlation functions for nearest- and third-nearest-neighbors are shown in Fig.~6. The data were obtained by the exact diagonalization in the N=20 lattice and by Mori's
technique in a $216 \times 216 $ lattice. In panels (a) and (b) ED correlations are constant and equal to unity in the ferromagnetic phase for $p < p_{20}.$
In this region correlation functions obtained by Mori's method are also constant. However, they are somewhat smaller than one due to approximations made in the S=1 case
[2]. The interaction between nearest neighbor spins vanishes at $p=1$, which manifests itself in the vanishing correlation $\langle \bf{S}_{\bf{0}}   \bf{S}_{\bf{a}} \rangle$. The correlation $\langle \bf{S}_{\bf{0}}   \bf{S}_{\bf{2a}} \rangle$
depends only weakly on $p$ in the range $p > p_{20}$ and $p > p_{cr}$. Analyzing condensation parameters, in Ref.[2] it was shown that at $T=0$ in the range $p_{cr} < p \lesssim 0.31$
there exists a spin-disordered phase. For larger frustration parameters the system becomes an antiferromagnet with an incommensurate ordering vector, which varies with $p$. Notice
that the phase transition at $p \approx 0.31$ does not reveal itself in Fig.~6. As seen from the figure, curves obtained in the larger lattice are more smooth than those in the
N=20 lattice, which, at least partly, may be connected with finite-size effects. However, spin correlations calculated in the two lattices by two different methods are in general
close and behave similarly with changing the frustration parameter. We can conclude that results obtained with the approximate approach based on Mori's projection operator
technique are in reasonable agreement with the exact-diagonalization data.

In conclusion, we  investigeted the spin-1 Heisenberg model on a triangular lattice with the ferromagnetic nearest- and antiferromagnetic third-nearest-neighbor exchange interactions with the use of the exact diagonalization of small lattices with periodic boundary conditions. The SPINPACK code using the Lanczos method was employed to find the energies of the ground and low-lying exited states in the entire range of the frustration parameter $0 < p < 1$, where $p = \frac{J_3}{J_3 - J_1}$, $J_1$ and  $J_3$ are the nearest- and third-nearest exchange constants.  Besides,  spin-spin correlation functions between nearest- and third-nearest spins and spin gaps were calculated.
We found qualitative and in some cases quantitative agreement between results on the ground-state energy and spin correlations, obtained by exact diagonalization of small lattices and by Mori's projection technique in larger lattices.


\section*{Acknowledgements}

This work was supported by the Estonian research project IUT2-27, the European Union through the
European Regional Development Fund
(Centre of Excellence "Mesosystems: Theory and Applications", TK114) and the ESF Grant No. 9371. P. R. thanks Prof. J. Richter for  the help in using  SPINPACK.

\newpage

Figure captions

Fig.~1. The triangular lattices with N=16 and N=20 sites, studied in this work.

Fig.~2. a) The dependencies of the ground state energy (E$_{GS}$, S=N=16, squares, solid line) and the energy of the lowest excited state with S=N-1=15 (E$_1$, circles, dashed line)  in the range of the frustration parameter $p \leq 0.45$ on the N=16 lattice. In this interval of $p$, the system is characterized by the ferromagnetic order, b) The dependencies of the ground state energy (E$_{GS}$, S=0, squares, solid line), the energy of the lowest singlet excited state  (E$_2$, circles, short-dashed line) and the energy of the lowest triplet excited state  (E$_3$, S=1, triangles,  dashed line) in the frustration parameter range $p \geq 0.45$.

Fig.~3. a) The dependencies of the ground state energy (E$_{GS}$, S=N=20, squares, solid line) and the energy of the lowest excited state with S=N-1=19 (E$_1$, circles, dashed line)  in the range of the frustration parameter $p \leq 0.28$ on the N=20 lattice. In this interval of $p$, the system is characterized by the ferromagnetic order, b) The dependencies of the ground state energy (E$_{GS}$, S=0, squares, solid line), the energy of the lowest singlet excited state  (E$_2$, circles, short-dashed line) and the energy of the lowest triplet excited state (E$_3$, S=1, triangles,  dashed line) in the frustration parameter range $p \geq 0.28$.

Fig.~4.  The dependencies of the spin gap  E$_{3}$(S=1) - E$_{GS}$(S=0)  on the frustration parameter $p$ for the N=16 (solid line) and  the N=20  (dashed line) lattices.

Fig.~5.  The dependencies of the ground state energy per site (E$_{GS}$/N)  on the frustration parameter $p$ for the N=20 lattice (solid line),  the N=16 lattice (dash-dotted line), obtained by ED, and in a $216 \times 216$  lattice, obtained by Mori's  projection operator technique (dashed line). The  parameters $p_{20}$, $p_{16}$ and $p_{cr}$ are values of  $p$, at which the transition from the ferromagnetic to the singlet GS occurs.

Fig.~6. . The nearest-neighbor (a) and third-nearest-neighbor spin correlations, obtained in the  N=20 lattice (circles and dashed lines) and in a $216 \times 216$ lattice
at $T=0$ by Mori's technique (squares and solid lines).


\begin{thebibliography}{}
\bibitem{SNakatsuju}S.~Nakatsuji, Y.~Nambu, H.~Tonomura, O.~Sakai, S.~Jonas, C.~Broholm,
H.~Tsunetsugu, Y.~Qiu, Y.~Maeno, Science 309 (2005) 1697.
\bibitem{r1} P.~Rubin, A.~Sherman, M.~Schreiber, Phys.\ Lett.\ A 376 (2012) 1062.
\bibitem{r2} P.~Rubin, A.~Sherman, M.~Schreiber, Acta Physica Polonica A 126 (2014) 242.
\bibitem{Mori}H.~Mori, Prog.\ Theor.\ Phys.\ 34 (1965) 399.
\bibitem{spin} J. Schulenburg, program package SPINPACK,
http://www-e.uni-magdeburg.de/jschulen/spin/.

\end{thebibliography}
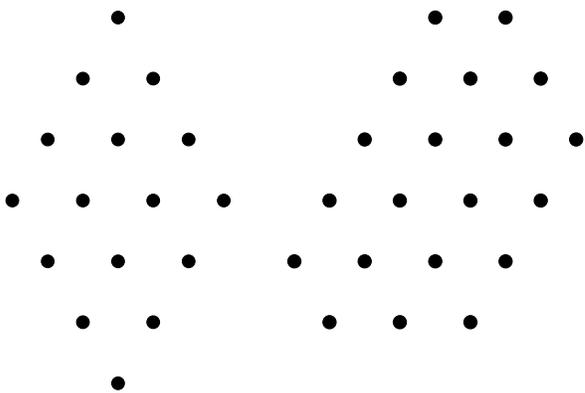
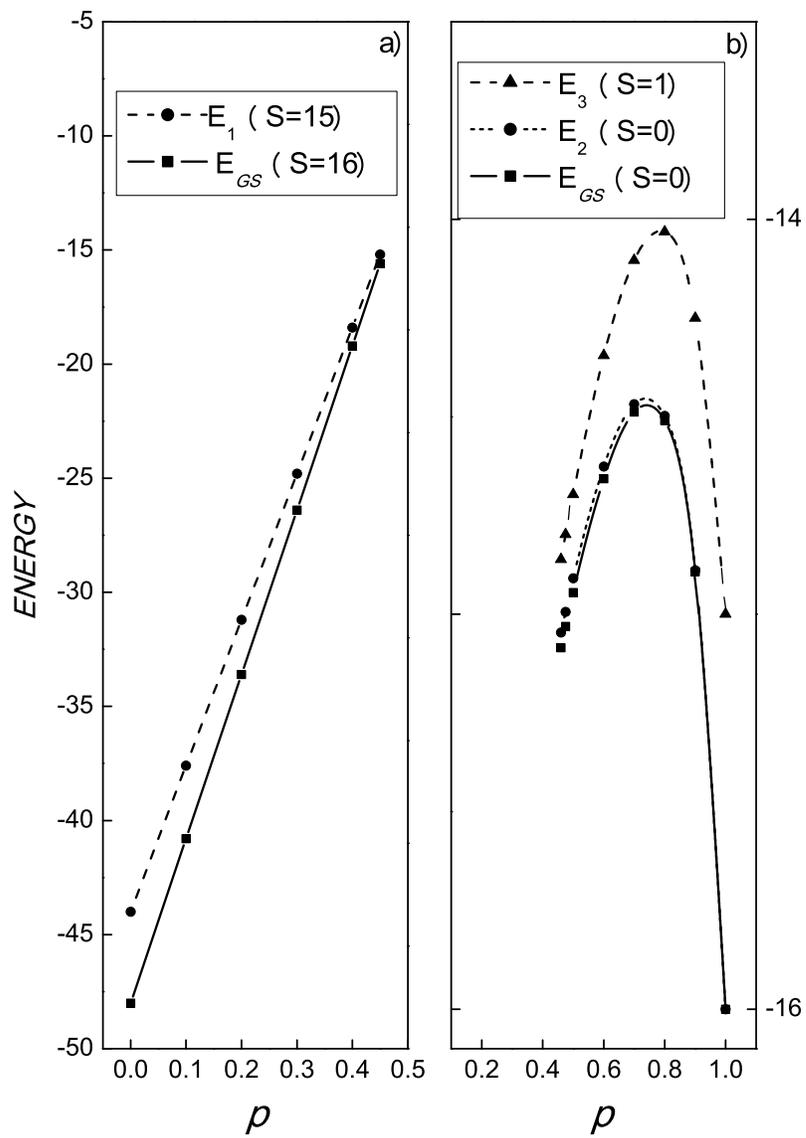
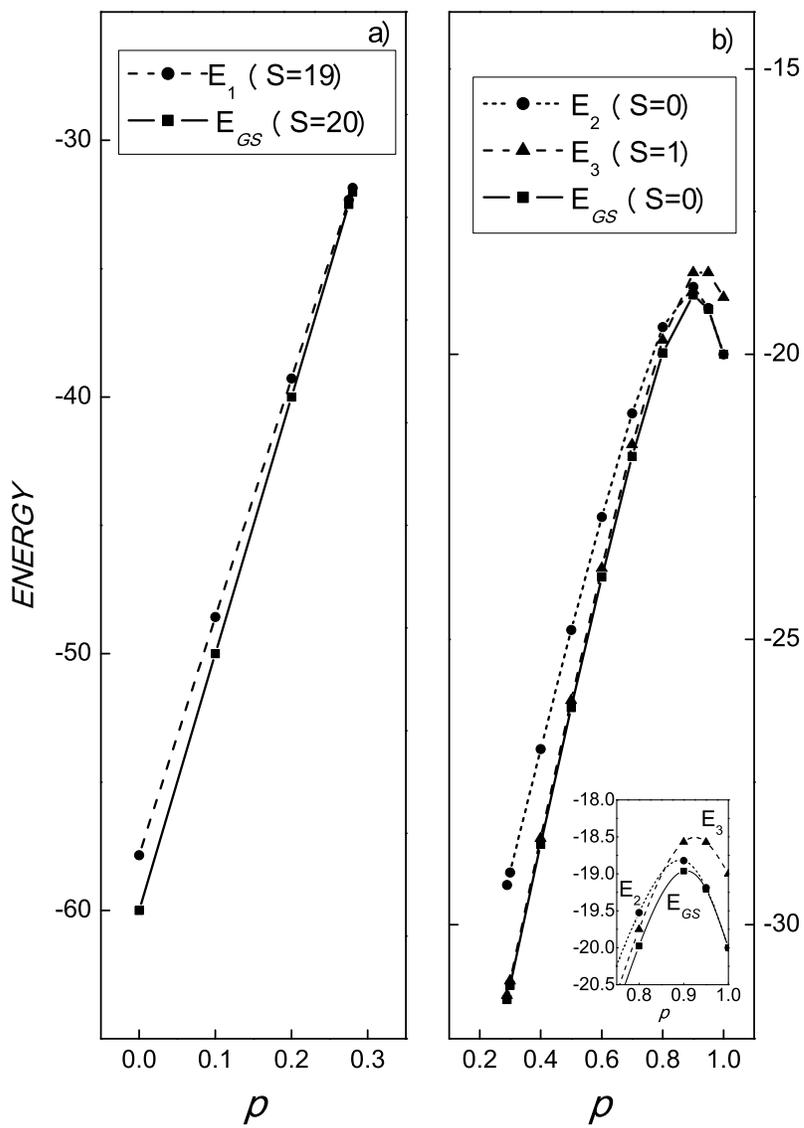
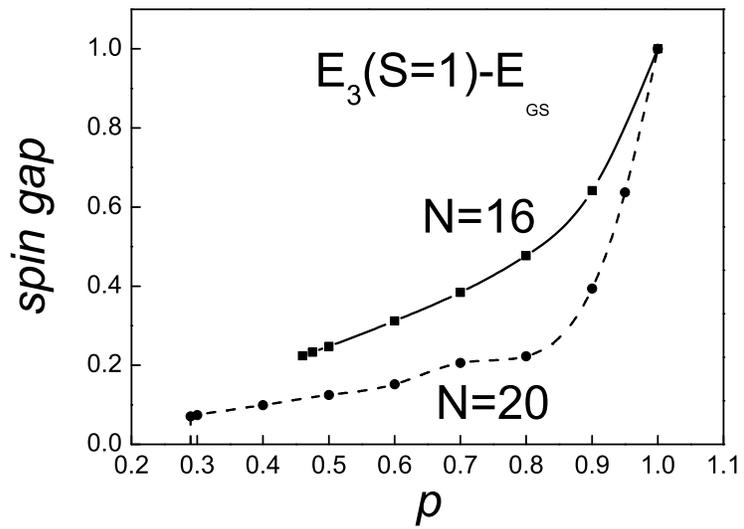
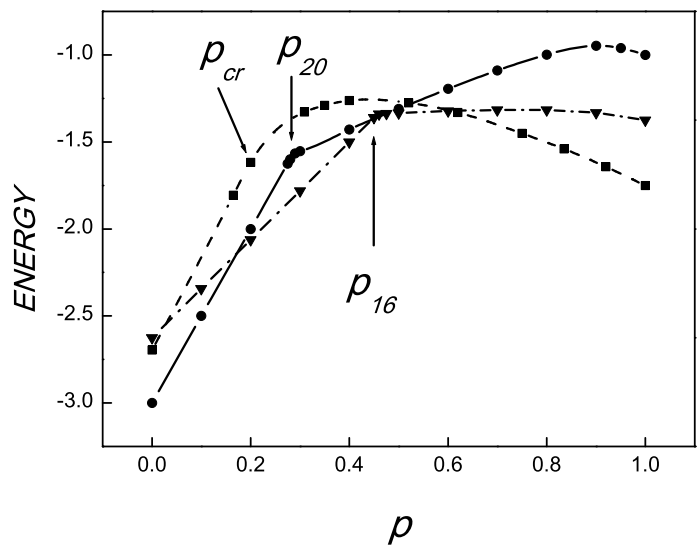
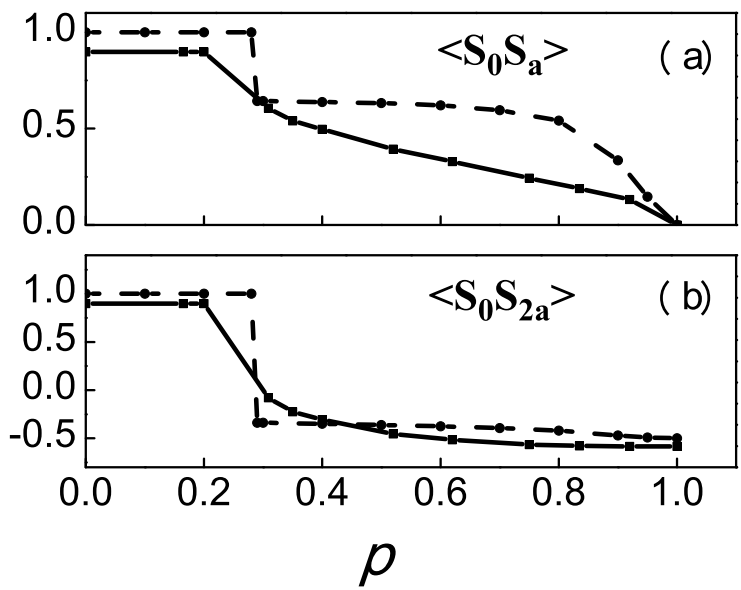
\end{document}